\documentclass[english,showpacs]{revtex4}
\usepackage[T1]{fontenc}
\usepackage[latin9]{inputenc}
\usepackage{amssymb}
\usepackage{graphicx}

\makeatletter

\usepackage{babel}
\makeatother

\begin{document}

\title{The Large-$N$ Limit of $PT$-Symmetric $O(N)$ Models}

\author{Hiromichi Nishimura and Michael Ogilvie}

\affiliation{Department of Physics, Washington University, St. Louis, MO, USA, 63130}

\date{October 20th, 2008}

\begin{abstract}
We study a $PT$-symmetric quantum mechanical model
with an $O(N)$-symmetric potential of the form $m^{2}\vec{x}^{2}/2-g(\vec{x}^{2})^{2}/N$
using its equivalent Hermitian form. Although the corresponding classical
model has finite-energy trajectories that escape to infinity, the
spectrum of the quantum theory is proven to consist only of bound
states for all $N$. We show that the model has two distinct phases
in the large-$N$ limit, with different scaling behaviors as $N$
goes to infinity. The two phases are separated by a first-order phase
transition at a critical value of the dimensionless parameter $m^{2}/g^{2/3}$,
given by $3\cdot2^{1/3}$.
\end{abstract}

\maketitle

\section{Introduction}

Models with $PT$ symmetry have emerged as an interesting extension
of conventional quantum mechanics.
There is a large class of models that are not Hermitian, but nevertheless have real spectra
as a consequence of $PT$-symmetry.
Bender and Boettcher
have shown
that single-component quantum mechanical models with $PT$-symmetric
potentials of the form $-\lambda\left(-ix\right)^{p}$ have real spectra
\cite{Bender:1998ke}.
An extensive literature on $PT$-symmetry and related matters now
exists, and there are extensive review articles available
\cite{Bender:2005tb,Bender:2007nj}.
Most of the results have been for models without continuous
internal symmetries, but there have been some results on models
with $O(N)$ symmetry 
\cite{Andrianov:1981wu,Ogilvie:2008tu,Andrianov:2007vt}. 
These models are particularly interesting
in the large-$N$ limit.
Meisinger and Ogilvie have shown that a $PT$-symmetric
version of the $O\left(N\right)$-invariant anharmonic oscillator
is isospectral with a Hermitian model with an $O(N-1)$ symmetry
 \cite{Ogilvie:2008tu}.
We study here the properties of this model, using its Hermitian form.

The Euclidean Lagrangian of the $PT$-symmetric model with $O(N)$ symmetry
is given by\begin{equation}
L_{E}=\sum_{j=1}^{N}\left[\frac{1}{2}\left(\partial_{t}x_{j}\right)^{2}+\frac{1}{2}m^{2}x_{j}^{2}\right]-\frac{g}{N}\left(\sum_{j=1}^{N}x_{j}^{2}\right)^{2}
\label{eq:EuclideanLagrangian}\end{equation}
where $g$ is positive. 
The minus sign in front of $g$ would lead to
a Hamiltonian unbounded from below if the model
were Hermitian.
From the standpoint of $PT$ symmetry,
the interaction term can be considered as a member of a family of
$PT$-invariant interactions\begin{equation}
-\frac{g}{N}\left(-\sum_{j=1}^{N}x_{j}^{2}\right)^{p}\end{equation}
which are invariant under $PT$ symmetry \cite{Bender:1998ke}. This
class of models is well-defined for $p=1$, and must be defined for
$p>1$ by an appropriate analytic continuation of the $x_{j}$ as
necessary. In \cite{Ogilvie:2008tu}, it was shown that this $PT$-symmetric model is
equivalent to a Hermitian model with Euclidean Lagrangian given by
\begin{equation}
L_{E}=\frac{1}{2}\dot{\sigma}^{2}+\frac{1}{2}\dot{\vec{\pi}}^{2}-m^{2}\sigma^{2}+\frac{4g}{N}\sigma^{4}+\frac{16g}{N}\sigma^{2}\vec{\pi}^{2}-\sqrt{2gN}\sigma\end{equation}
 where $\sigma$ is a single variable, and $\vec{\pi}$ is a vector of
$N-1$ variables. The corresponding Hamiltonian $H$ is
\begin{equation}
H=-\frac{1}{2}\frac{\partial^{2}}{\partial\sigma^{2}}-\frac{1}{2}\frac{\partial^{2}}{\partial\vec{\pi}^{2}}-m^{2}\sigma^{2}+\frac{4g}{N}\sigma^{4}+\frac{16g}{N}\sigma^{2}\vec{\pi}^{2}-\sqrt{2gN}\sigma.
\end{equation}
This Hermitian form has many remarkable features.
First of all, it has a manifest $O(N-1)$ symmetry associated with
rotations of $\vec{\pi}$ rather than the $O(N)$ symmetry
of equation~(\ref{eq:EuclideanLagrangian}). As in the similar case of 
a $PT$-symmetric $-gx^{4}$ theory
of a single variable \cite{Bender:2006wt,Jones:2006et}, 
there is a linear anomaly term which breaks
the classical symmetry $\sigma\rightarrow-\sigma$ at order $\hbar$.
The $\vec\pi$ field has no quadratic mass term, while the sign
of the mass term for the $\sigma$ field is opposite the sign of the
$x$ fields in the original Lagrangian. From a naive field-theoretic
point of view, the $\vec{\pi}$ field is massless at tree level. 

There are two important questions we will address. The first question
is the presence or absence of scattering states in this model for
any value of $N$. It is clear that the corresponding classical model
has a class of trajectories with finite, continuously varying energies
that escape to infinity along paths with $\sigma=0$. Nevertheless,
we will prove that the quantum mechanical model has only bound states
with discrete energy levels. This behavior is similar to that of the
quantum-mechanical $x^{2}y^{2}$ model.
This model can be derived from the quantum-mechanical reduction
of a two-dimensional gauge theory \cite{Matinyan:1981dj} and was originally thought to be fully ergodic, \textit{i.e.}, to have only chaotic motion. 
This turns out
not to be the case \cite{Dahlqvist:1990zz}. 
Simon has shown that the quantum mechanical
version of the $x^2y^2$ model has a purely discrete spectrum
\cite{Simon:1983jy}.
As we will show in section \ref{sec:Simon}, the arguments
of Simon can be generalized to show
that the $PT$-symmetric $O(N)$ models have discrete spectrum for all finite values of $N$. 

The second, more difficult question, concerns the existence and interpretation
of the large-$N$ limit. We explain the nature of the problem in section
III. In section IV, we show that a simple variational approximation
gives us the clues we need to prove that this model has two distinct
scaling behaviors in the large-$N$ limit, controlled by the
dimensionless parameter $m^{2}/g^{2/3}$.
A first-order transition occurs in the large-$N$ limit
when $m^{2}/g^{2/3}=3\cdot2^{1/3}$.
Because this is a quantum mechanical system, the phase transition only
appears in the strict limit of $N\rightarrow\infty$. For large but
finite $N,$ there is a rapid crossover between the two different
scaling behaviors. Section V rederives the results of section IV
using the Born-Oppenheimer approximation, which proves to be
exact in the large-$N$ limit.
In section VI we present the results of a numerical study of the ground
state energy for $m^{2}=0$ that confirm our analytical results.
A final section presents our conclusions.

\section{Absence of Scattering States}\label{sec:Simon}

We now turn to the issue of the spectrum for finite $N$.
In \cite{Simon:1983jy}, Simon gave five arguments for the absence of scattering states
in an $x^{2}y^{2}$ potential model. His first, and simplest, argument
is based on a lower bound for the Hamiltonian\begin{equation}
H_{xy}=-\frac{\partial^{2}}{\partial x^{2}}-\frac{\partial^{2}}{\partial y^{2}}+x^{2}y^{2}\end{equation}
using the operator inequality\begin{equation}
-\frac{\partial^{2}}{\partial x^{2}}+x^{2}y^{2}\ge|y|\end{equation}
 which obviously holds on a harmonic oscillator basis. Applying this
inequality symmetrically to $x$ and $y$, we see that\begin{equation}
H_{xy}\ge \frac{1}{2}\left[-\frac{\partial^{2}}{\partial x^{2}}-\frac{\partial^{2}}{\partial y^{2}}+\left|x\right|+\left|y\right|\right].\end{equation}
This lower bound Hamiltonian has only bound states.
This key step then leads via the Golden-Thompson inequality
to the conclusion that $H_{xy}$ has purely discrete spectrum.

Applying the same ideas to our Hamiltonian, we see that\begin{equation}
H\ge-\frac{1}{4}\frac{\partial^{2}}{\partial\sigma^{2}}-\frac{1}{4}\frac{\partial^{2}}{\partial\vec{\pi}^{2}}-m^{2}\sigma^{2}+\frac{4g}{N}\sigma^{4}-\sqrt{2gN}\sigma
+\sqrt{\frac{2g}{N\left(N-1\right)}}\sum_{j=1}^{N-1}\left|\pi_{j}\right|
+\left(N-1\right)\sqrt{\frac{2g}{N}}\left|\sigma\right|.\end{equation}
 The specific bounding Hamiltonian is not so important; the crucial
feature is that the bounding Hamiltonian has only bound states. This
is sufficient to guarantee that $H$ has only bound states for all
finite values of $N$.

\section{Scaling Arguments and the Large-$N$ Limit}

\begin{figure}
\includegraphics{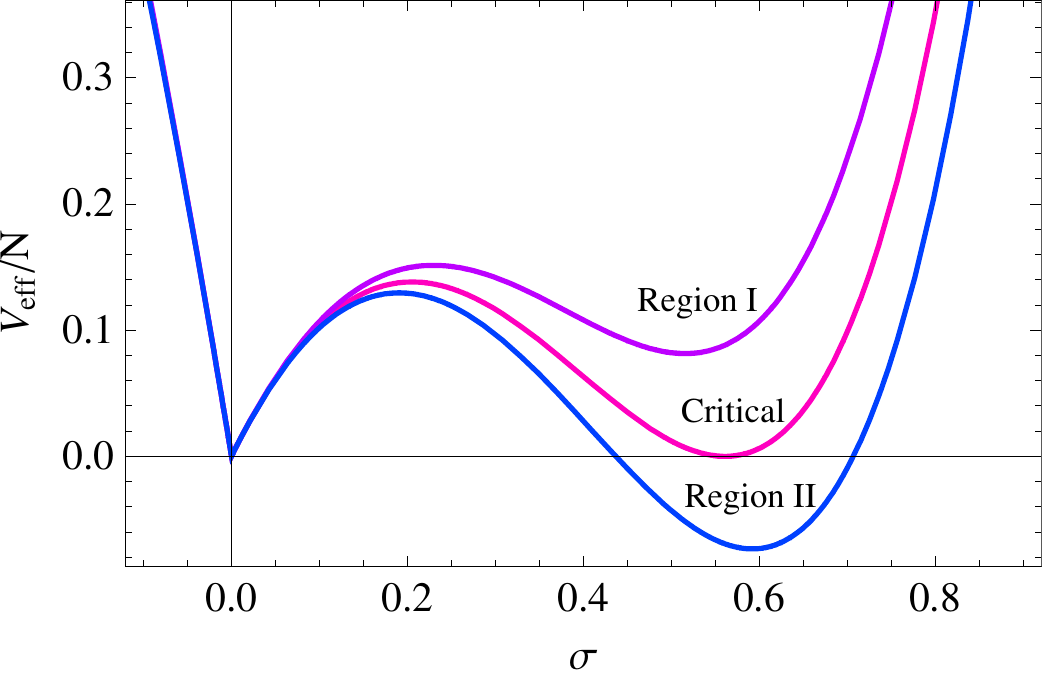}
\caption{The effective potential $V_{eff}/N$ versus $\sigma$ for $g=1$. Three different values of $m^2$ are shown, corresponding to region I, region II, and the critical point.}
\label{fig:Veff}
\end{figure}

We now consider the large-$N$ limit of the model. Our naive expectation
based on the Hermitian $O(N)$ model is that the ground state energy
will be proportional to $N$ as $N\rightarrow\infty$ 
\cite{Coleman:1974jh}. 
In order to explore this possibility, we rescale
the Lagrangian $L_E$  by $\sigma\rightarrow\sqrt{N}\sigma$ to obtain
\begin{equation}
L_{E}=\frac{N}{2}\dot{\sigma}^{2}+\frac{1}{2}\dot{\vec{\pi}}^{2}-Nm^{2}\sigma^{2}+4gN\sigma^{4}+16g\sigma^{2}\vec{\pi}^{2}-N\sqrt{2g}\sigma.\end{equation}
 We see that the anomaly term survives in the large-$N$ limit, unlike
the $PT$-symmetric matrix case \cite{Ogilvie:2008tu}. 
After integrating over the
$N-1$ $\vec{\pi}$ fields, we have a large-$N$ effective potential
$V_{eff}$ for $\sigma$:\begin{equation}
V_{eff}/N=-m^{2}\sigma^{2}+4g\sigma^{4}+\frac{1}{2}\sqrt{32g\sigma^{2}}-\sqrt{2g}\sigma.\end{equation}
The new term $\frac{1}{2}\sqrt{32g\sigma^{2}}$ comes from the functional
determinant for fluctuations of the $\vec{\pi}$ fields, and represents
the zero-point energy of their quantum fluctuations given a constant
value for $\sigma$.
The shape of the potential is controlled by the dimensionless parameter
$m^2/g^{2/3}$.
Figure~\ref{fig:Veff}  shows the effective potential as a function of $\sigma$
 with $g$ set to $1$ for three different values of $m^2$.
We refer to
the region where $m^{2}/g^{2/3}<3\cdot2^{1/3}$ as region I, and the
region where $m^{2}/g^{2/3}>3\cdot2^{1/3}$ as region II.
In region I, $V_{eff}$ has a global minimum at $\sigma=0$.
For $\sigma$ near $0$, $V_{eff}$ is approximately linear,
but with different slopes for $\sigma>0$ and $\sigma<0$.
This occurs because the zero-point energy of the $\vec{\pi}$ fields
has virtually the same form as the anomaly term. However,
the zero-point energy term respects a discrete, classical
 $\sigma\rightarrow-\sigma$
symmetry which the anomaly explicitly breaks.
The boundary between regions I and II is given by
$m^{2}/g^{2/3}=3\cdot2^{1/3}$, where $V_{eff}$ has two
degenerate minima. 
In region II, $V_{eff}$ has a global minimum with
$\sigma\ne 0$.
This
change in the behavior of the effective potential as $m^{2}$ is varied
is not seen in the corresponding Hermitian model \cite{Coleman:1974jh}, 
and naively indicates that the $N-1$ $\vec{\pi}$ fields
are massless modes in region I. 

To understand better the suspect character of the above analysis, it is useful
to rederive these results using hyperspherical coordinates  \cite{Witten:1980xxx,Chatterjee:1990se}. This formalism will also be used later in the context
of the Born-Oppenheimer approximation. If we define $\rho=\left(\vec{\pi}^{2}\right)^{1/2}$, 
the reduced, radial Hamiltonian in the sector of angular momentum $l$ is
\begin{equation}
H_{rad}=-\frac{1}{2}\frac{\partial^{2}}{\partial\sigma^{2}}-\frac{1}{2}\frac{\partial^{2}}{\partial\rho^{2}}+\frac{\left(N+2l-3\right)\left(N+2l-1\right)}{8\rho^{2}}-m^{2}\sigma^{2}+\frac{4g}{N}\sigma^{4}+\frac{16g}{N}\sigma^{2}\rho^{2}-\sqrt{2gN}\sigma
\label{eq:radialH}\end{equation}
 where $l$ is a non-negative integer. After rescaling $\sigma\rightarrow\sqrt{N}\sigma$
and $\rho\rightarrow\sqrt{N}\rho$, we have\begin{equation}
H_{rad}=-\frac{1}{2N}\frac{\partial^{2}}{\partial\sigma^{2}}-\frac{1}{2N}\frac{\partial^{2}}{\partial\rho^{2}}+\frac{\left(N+2l-3\right)\left(N+2l-1\right)}{8N\rho^{2}}-Nm^{2}\sigma^{2}+4gN\sigma^{4}+16gN\sigma^{2}\rho^{2}-N\sqrt{2g}\sigma\end{equation}
Taking $l=0$, we see that the ground state energy to leading order
in large-$N$ is given by minimizing\begin{equation}
N\left[\frac{1}{8\rho^{2}}-m^{2}\sigma^{2}+4g\sigma^{4}+16g\sigma^{2}\rho^{2}-\sqrt{2g}\sigma\right]\end{equation}
with respect to $\rho$ and $\sigma$. Minimizing with respect to
$\rho$, we find\begin{equation}
\rho=\left(128g\sigma^{2}\right)^{-1/4}\end{equation}
 and must now minimize\begin{equation}
N\left[-m^{2}\sigma^{2}+4g\sigma^{4}+2\sqrt{2g\sigma^{2}}-\sqrt{2g}\sigma\right]\end{equation}
which is identical to our previous expression for $V_{eff}$. However,
we now notice that when $\sigma=0$, $\rho$ is infinite. This strongly
suggests that our treatment of the large-$N$ limit is not valid in region I.

\section{A Simple Variational Approximation}

In order to gain analytical insight about the large-$N$ behavior
of this model, we apply a simple variational approximation using harmonic
oscillator ground states. We assume that the ground state has the
form\begin{equation}
\Psi\left[\sigma,\vec{\pi}\right]=\psi_{0}\left(\sigma\right)\prod_{j=1}^{N-1}\phi_{0}\left(\pi_{j}\right)\end{equation}
where $\phi_{0}$ is a harmonic oscillator ground state of frequency
$\Omega$ and expected value $\left\langle \pi_{j}\right\rangle =0$.
The wave function $\psi_{0}\left(\sigma\right)$ has frequency $\omega$
and expected value $\left\langle \sigma\right\rangle =v$. We then
have the variational inequality $E_{0}\le E_{var}$ where\begin{equation}
E_{var}=\frac{1}{4}\omega+\frac{N-1}{4}\Omega-m^{2}\left(v^{2}+\frac{1}{2\omega}\right)+\frac{4g}{N}\left(v^{4}+\frac{3v^{2}}{\omega}+\frac{3}{4\omega^{2}}\right)+\frac{16g}{N}\left(v^{2}+\frac{1}{2\omega}\right)\frac{N-1}{2\Omega}-\sqrt{2gN}v\end{equation}
provides an upper bound for all $v$ and positive $\omega$ and $\Omega$.
We are free to rescale the variational parameters $v$, $\omega$,
and $\Omega$ as we like, and we have considered the class of rescalings
of the form $v\rightarrow N^{a}v$, $\omega\rightarrow N^{b}\omega$
, and $\Omega\rightarrow N^{c}\Omega$. There are two different rescalings
with non-trivial behavior in the large-$N$ limit.

If we rescale the variational parameters as $v\rightarrow N^{1/2}v$,
$\omega\rightarrow N\omega$ , leaving $\Omega$ unchanged, we find
after some algebra that the ground state energy in this large-$N$
limit is given by minimizing \begin{equation}
E_{var}=N\left[\frac{\omega}{4}+\frac{\Omega}{4}-m^{2}v^{2}+4gv^{4}+\frac{8gv^{2}}{\Omega}-\sqrt{2g}v\right].\end{equation}
This in turn reduces to minimizing
\begin{equation}
E_{var}=N\left[-m^{2}v^{2}+4gv^{4}+2\sqrt{2gv^{2}}-\sqrt{2g}v\right]\end{equation}
 with respect to $v$. Unsurprisingly, this is equivalent to minimizing
our previous expression for the effective potential in the conventional
large-$N$ limit and is valid in region II.

A different scaling behavior, which we will show is valid in region I
in the next section, is obtained
if we perform rescalings $v\rightarrow N^{-1/6}v$, $\omega\rightarrow N^{1/3}\omega$, and $\Omega\rightarrow N^{-2/3}\Omega$. 
This yields a large-$N$
limit of the form\begin{equation}
E_{var}=N^{1/3}\left[\frac{\omega}{4}+\frac{\Omega}{4}+\left(v^{2}+\frac{1}{2\omega}\right)\frac{8g}{\Omega}-\sqrt{2g}v\right].\end{equation}
Minimization of $E_{var}$ with respect to the three parameters $\omega,$
$\Omega$, and $v$ leads to the solution
\begin{eqnarray}
\Omega&=&\left(\frac{16}{3}\right)^{2/3}g^{1/3}\\
\omega&=&4\left(\frac{16}{3}\right)^{-1/3}g^{1/3}\\
v&=&2^{-5/6}3^{-2/3}g^{-1/6}\\
E_{var}&=&\left(\frac{3}{2}\right)^{4/3}N^{1/3}g^{1/3}.\end{eqnarray}
 Numerically, this gives an upper bound on $E_{0}$ of approximately
$1.71707\, N^{1/3}g^{1/3}$. As we show below, this variational result
is exact in the large-$N$ limit in region I.

\section{The Born-Oppenheimer Approximation and the Large-$N$ Limit}

We now apply the two different scaling behaviors we have found in
the previous section to the Hamiltonian directly. If the $\sigma$
field is rescaled by $\sigma\rightarrow N^{1/2}\sigma$, the rescaled
Hamiltonian is\begin{equation}
H=-\frac{1}{2N}\frac{\partial^{2}}{\partial\sigma^{2}}-\frac{1}{2}\frac{\partial^{2}}{\partial\vec{\pi}^{2}}-Nm^{2}\sigma^{2}+4gN\sigma^{4}+16g\sigma^{2}\vec{\pi}^{2}-N\sqrt{2g}\sigma\end{equation}
which will turn out to be valid in region II. All of the terms
in the Hamiltonian are of order $N$, except for the kinetic energy
term for $\sigma$, which is of order $1/N$. This suggests the use
of the Born-Oppenheimer approximation,
in which heavy degrees of freedom are treated classically. 
In this approximation, the
wave function is written as $\psi_{II}\left(\vec{\pi},\sigma\right)=u_{II}\left(\vec{\pi},\sigma\right)w_{II}\left(\sigma\right)$
where $u_{II}\left(\vec{\pi},\sigma\right)$ satisfies\begin{equation}
\left[-\frac{1}{2}\frac{\partial^{2}}{\partial\vec{\pi}^{2}}+16g\sigma^{2}\vec{\pi}^{2}\right]u_{II}\left(\vec{\pi},\sigma\right)=\epsilon_{II}\left(\sigma\right)u_{II}\left(\vec{\pi},\sigma\right),\end{equation}
describing the $N-1$ $\vec{\pi}$ fields as harmonic oscillators
with frequencies determined by $\sigma.$ The total energy to leading
order in $N$ is again the effective potential $V_{eff}$. As we have
seen, in region II there is a non-trivial solution for which the rescaled
ground state energy $E_{0}/N$ has a finite, negative value as $N\rightarrow\infty$
. In region I, we find that $E_{0}/N\rightarrow0$ as $N\rightarrow\infty$
; the vacuum expectation value of the original, un-rescaled $\sigma$
also obeys $N^{-1/2}\left\langle \sigma\right\rangle \rightarrow0$
in this limit as well. 

The other rescaling is $\sigma\rightarrow N^{-1/6}\sigma$ combined
with $\vec{\pi}\rightarrow N^{5/6}\vec{\pi}.$ The rescaled Hamiltonian
is\begin{equation}
H=N^{1/3}\left[-\frac{1}{2}\frac{\partial^{2}}{\partial\sigma^{2}}-\frac{1}{2N^{2}}\frac{\partial^{2}}{\partial\vec{\pi}^{2}}+16g\sigma^{2}\vec{\pi}^{2}-\sqrt{2g}\sigma-N^{-2/3}m^{2}\sigma^{2}+4gN^{-2}\sigma^{4}\right].\end{equation}
With this rescaling, it appears energies will increase with $N$ as $N^{1/3}$.
We also see from the kinetic energy term for $\sigma$ that the $\sigma$
field retains its operator character as $N$ goes to infinity, unlike
the conventional large-$N$ limit obtained in region II. We can immediately
drop the $\sigma^{2}$ and $\sigma^{4}$ terms as irrelevant in the
large-$N$ limit. The relevance of the $\vec{\pi}$ kinetic field
is less clear. If we again use hyperspherical coordinates for $\vec{\pi}$,
we find that the effective Hamiltonian in the $l=0$ sector can be
written for large $N$ as\begin{equation}
H_{eff}=N^{1/3}\left[-\frac{1}{2}\frac{\partial^{2}}{\partial\sigma^{2}}-\frac{1}{2N^{2}}\frac{\partial^{2}}{\partial\rho^{2}}+\frac{1}{8\rho^{2}}+16g\sigma^{2}\rho^{2}-\sqrt{2g}\sigma\right]\end{equation}
 where $\rho$ is now the magnitude of the rescaled $\vec{\pi}$.
It is clear that $\rho$ has a mass of order $N^{2}$, and we
again use the Born-Oppenheimer approximation, this time applied to
$\rho$. The wave function is written as 
$\psi_{I}\left(\vec{\pi},\sigma\right)=u_{I}\left(\sigma,\rho\right)w_{I}\left(\rho\right)$
where $u_{I}\left(\sigma,\rho\right)$ satisfies\begin{equation}
\left[-\frac{1}{2}\frac{\partial^{2}}{\partial\sigma^{2}}+16g\sigma^{2}\rho^{2}-\sqrt{2g}\sigma\right]u_{I}\left(\sigma,\rho\right)=\epsilon_{I}\left(\rho\right)u_{I}\left(\sigma,\rho\right),\end{equation}
 which describes a particle in a harmonic potential. The ground state
energy is $2\sqrt{2g}\rho-1/\left(32\rho^{2}\right)$. The energy of
the combined system is given by minimizing\begin{equation}
\epsilon_{I}\left(\rho\right)+\frac{1}{8\rho^{2}}=\frac{3}{32\rho^{2}}+2\sqrt{2g}\rho\end{equation}
 which gives the ground state energy in the large-$N$ limit in region
I to be exactly given by\begin{equation}
E_{0}^{\left(I\right)}=\left(\frac{3}{2}\right)^{4/3}N^{1/3}g^{1/3}\end{equation}
 which is identical to the result of the variational treatment in section IV.

We now have results for two different large-$N$ limits. In one case,
the ground state energy is proportional to $N$ \begin{equation}
E_{0}^{\left(II\right)}=N\cdot\min_{\sigma}\left[-m^{2}\sigma^{2}+4g\sigma^{4}+\frac{1}{2}\sqrt{32g\sigma^{2}}-\sqrt{2g}\sigma\right].\end{equation}
 In the other case, the ground state energy \begin{equation}
E_{0}^{\left(I\right)}=\left(\frac{3}{2}\right)^{4/3}N^{1/3}g^{1/3}\end{equation}
 is proportional to $N^{1/3}$ and is always positive. If $E_{0}^{\left(II\right)}$
is negative, as it is in region II, it will be favored over $E_{0}^{\left(I\right)}$.
In region I, the formula for $E_{0}^{\left(II\right)}$ appears to
predict that the ground state energy is zero. However, this is misleading:
it is actually predicting that the ground state energy is not growing
linearly with $N,$ but at some less rapid rate. This kind of behavior
is shown by $E_{0}^{\left(I\right)}$, and the two expressions are
in fact consistent. If we examine the behavior of the scaled ground
state energy $E_{0}/N$ in the large-$N$ limit, it is zero in region
I and negative in region II. However, the true behavior of $E_{0}$
in region I is given by $E_{0}^{\left(I\right)}$. Note that similar
considerations apply to the expectation value $\left\langle \sigma\right\rangle $.
In region I, $\left\langle \sigma\right\rangle $ is decreasing as
$N^{-1/6}$, while in region II it is increasing as $N^{1/2}$.

In the large-$N$ limit, these two behaviors are completely incommensurate,
and give rise to a first-order transition at $m^{2}/g^{2/3}=3\cdot2^{1/3}$,
the boundary between regions I and II. However, there can be no phase
transition in quantum mechanics for any finite number of degrees of
freedom. We can understand the behavior of these two solutions for
finite but large values of $N$ from the effective potential $V_{eff}$
in Figure \ref{fig:Veff}.
In the region of parameter space
where $m^{2}/g^{2/3}$ is close to $3\cdot2^{1/3}$, $V_{eff}$ has
two distinct local minima, and there will be tunneling between these
two minima, leading to a small splitting of the ground state energy from
that of the first excited state. This is very similar to the behavior of the
double well with a small term linear in the coordinate $x$ added. In this case,
however, as $N$ becomes large, the tunneling between the two minima
is suppressed, leading to a first-order transition in the large-$N$
limit.

\section{The Case $m^{2}=0$}

\begin{figure}
\includegraphics{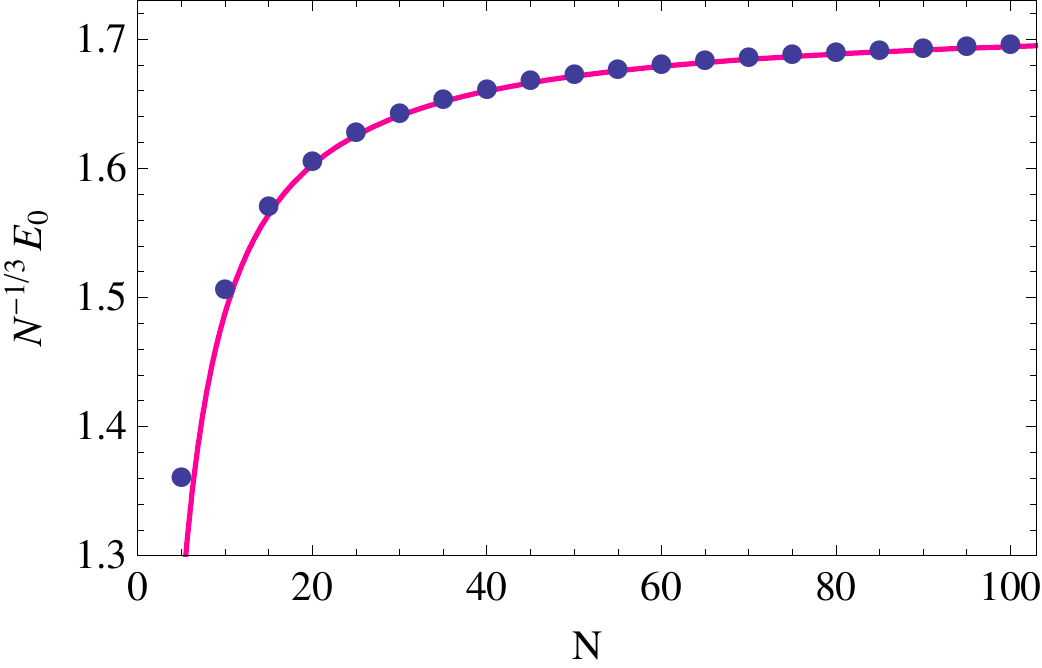}
\caption{The variational estimate of the scaled ground state energy versus $N$ at $m^2=0$.}
\label{fig:E0-vs-N}
\end{figure}

In order to check our large-$N$ result in region I,
we study numerically the scaling behavior of the ground state energy
for the case $m^{2}=0$ for large values of $N$.
This value for
$m^{2}$ is in region I, where the ground state energy scales as $N^{1/3}$
in the large-$N$ limit. As we have seen in the previous section,
the $m^{2}$ dependence
disappears in this limit in region I, so the case $m^{2}=0$ gives
the large-$N$ behavior throughout this region. 
After rescaling the fields
 by $\sigma\rightarrow g^{-1/6}\sigma$ and
$\vec{\pi}\rightarrow g^{-1/6}\vec{\pi}$,  
the radial Hamiltonian $H_{rad}$ is given by
\begin{eqnarray}
(gN)^{-1/3}H_{rad}&=&-\frac{1}{2}\frac{\partial^{2}}{\partial\sigma^{2}}+\frac{1}{N}\left(-\frac{1}{2}\frac{\partial^{2}}{\partial\rho^{2}}+\frac{\left(N+2l-3\right)\left(N+2l-1\right)}{8\rho^{2}}+16\sigma^{2}\rho^{2}\right)\nonumber\\
&&-\sqrt{2}\sigma-N^{-2/3}m^{2}g^{-2/3}\sigma^{2}+4N^{-2}\sigma^{4}.\end{eqnarray}
Clearly, $g$ can be set to $1$ at $m^2 = 0 $ with no loss of generality.
Different angular momentum sectors do not mix, and we restrict
ourselves to the $l=0$ sector. 

We have performed a series of variational calculations using a harmonic
oscillator basis for both $\sigma$ and $\rho$. In these calculations,
a finite-dimensional square matrix is formed from the matrix elements
of $H_{rad}$ using a suitable basis. The lowest eigenvalue of
this matrix is an upper bound on the ground state energy. For $\rho$,
these basis elements were chosen to be solutions of the reduced Hamiltonian
for the $\left(N-1\right)$-dimensional harmonic oscillator. With
the angular frequency scaled to $1$, the
general form of a basis function for $\rho$
is $\rho^{{N}/{2}-1+l}L_{\frac{m-l}{2}}^{\left(N-3\right)/2+l}(\rho^{2})\exp\left[-\frac{1}{2}\rho^{2}\right]$, 
where $m$ is a non-negative integer,
and has energy $m+\left(N-1\right)/2$. Basis sets of size $n^{2}$
were constructed using $n$ basis elements for both $\sigma$
and $\rho$ . The optimum values for the angular frequencies associated
with $\sigma$ and $\rho$ were determined by minimizing the lowest eigenvalue
using a nine-element basis. These angular frequencies were then used
with a 36-element basis to estimate the ground state energy. In Figure
\ref{fig:E0-vs-N}, we plot the 36-element estimate for the ground state energy,
scaled by a factor of $N^{-1/3}$ , versus $N$. We also show a fit
of the form\begin{equation}
N^{-1/3}E_{0}=a+\frac{b}{N}.\end{equation}
Although the fit was obtained using the points from $N=50$ to $N=100$,
the agreement with the calculated energies is quite
reasonable even at $N=5$.
The constants obtained from this fit are $a=1.71705(4)$ and
$b=-2.29(2)$, where the errors are estimated from
Richardson extrapolation and varying the fitting conditions.
These estimated errors do not represent the true systematic errors due,
{\it e.g.}, to truncation to a finite basis.
Note the excellent agreement of $a$ with the exact result 
$\left(\frac{3}{2}\right)^{4/3}\approx 1.71707$.

\section{Conclusions}

The $PT$-symmetric $O(N)$ model has been analyzed in a fairly
complete way using its dual Hermitian form.
The Hermitian form of the Hamiltonian has classical
trajectories of finite energy that escape to infinity,
and also  $N-1$ variables which are massless
in perturbation theory.
Nevertheless, the model can be proven to have only
discrete energy eigenstates.
There are two distinct
 large-$N$ limits, with
incommensurate
scaling behavior. 
For $m^{2}/g^{2/3} > 3\cdot2^{1/3}$, the large-$N$ limit
is similar in behavior to that of the Hermitian $O(N)$ model,
with a ground state energy depending on $m$ and proportional to $N$.
For $m^{2}/g^{2/3} < 3\cdot2^{1/3}$,
the ground state energy is proportional to $N^{1/3}$,
and is independent of $m$ throughout this region.
The two regions are separated by a first-order phase transition
in the limit $N\rightarrow\infty$.
All of these results were obtained from the dual Hermitian
form of the original $PT$-symmetric model.
Although it was shown in \cite{Ogilvie:2008tu} that the large-$N$ limit
of region II could be obtained from the $PT$-symmetric form,
these arguments do not appear to give the detailed information
for region I that can be obtained from the Hermitian form.
It would be very desirable to achieve an understanding
of this model for both regions using only the $PT$-symmetric form
of the model. Such an understanding might be an important step
in constructing $PT$-symmetric scalar field theories.

\end{document}